\pdfoutput=1

\documentclass[11pt]{article}

\usepackage[preprint]{acl}

\usepackage{times}
\usepackage{latexsym}
\usepackage{enumitem}
\usepackage[T1]{fontenc}

\usepackage[utf8]{inputenc}

\usepackage{inconsolata}

\usepackage{graphicx}

\usepackage{url}      
\usepackage{todonotes} 
\usepackage{tcolorbox}

\newcommand{\systemname}{\textsc{SeaView}}{}

\DeclareRobustCommand{\seaview}{%
  \begingroup\normalfont
  \vspace{-0.2em}%
  \raisebox{-0.4em}{%
  \includegraphics[height=1.5em]{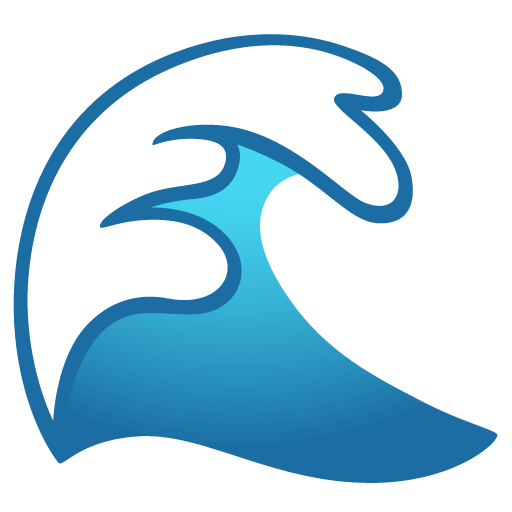}%
  }%
  \kern 0.4em%
  \endgroup
}

\title{\seaview \systemname{}: Software Engineering Agent Visual Interface for Enhanced Workflow}

\author{
Timothy Bula\thanks{Corresponding author}, Saurabh Pujar, Luca Buratti, Mihaela Bornea, Avirup Sil\\
\texttt{\{trbula, mabornea, avi\}@us.ibm.com}\\
\texttt{\{saurabh.pujar, luca.buratti1\}@ibm.com}\\
IBM Research AI
}

\begin{document}
\maketitle

\begin{abstract}
 Auto-regressive LLM-based software engineering (SWE) agents, henceforth SWE agents, have made tremendous progress (>60\% improvement on SWE-Bench) on real-world coding challenges including GitHub issue resolution. SWE agents use a combination of reasoning, environment interaction and self-reflection to resolve issues thereby generating ``trajectories''. Analysis of SWE agent trajectories is difficult, not only as they exceed LLM sequence length (sometimes, greater than 128k) but also because it involves a relatively prolonged interaction between an LLM and the environment managed by the agent.
In case of an agent error, it can be hard to decipher, locate and understand its scope. Similarly, it can be hard to track improvements or regression over multiple runs or experiments. While a lot of research has gone into making these SWE agents reach state-of-the-art, much less focus has been put into creating tools to help analyze and visualize agent output. We propose a novel tool called \systemname{}: \textbf{S}oftware \textbf{E}ngineering \textbf{A}gent \textbf{V}isual \textbf{I}nterface for \textbf{E}nhanced \textbf{W}orkflow, with a vision to assist SWE-agent researchers to visualize and inspect their experiments. \systemname{}'s novel mechanisms help compare experimental runs with varying hyper-parameters or LLMs, and quickly get an understanding of LLM or environment related problems. Based on our user study, experienced researchers spend between 10 and 30 minutes to gather the information provided by \systemname{}, while researchers with little experience can spend between 30 minutes to 1 hour to diagnose their experiment. 

\end{abstract}

\section{Introduction}
The potential of large language models (LLMs) to streamline software engineering (SWE) workflows is evident, particularly demonstrated by their enhanced performance on evaluation datasets such as SWE-Bench \cite{jimenez2024swebench} and Commit0 \cite{zhao2024commit0}. Although advancements in agent interaction methodologies \cite{yang2024swe} and prompt engineering \cite{wang-etal-2024-wordflow} have yielded substantial improvements in SWE agents, the visualization and interpretability of these agents' internal processes remain comparatively under-explored.
Current SWE agents \cite{wang2024openhands, antoniades2024swe, yang2024swe} combine the use of powerful LLMs \cite{achiam2023gpt, claude2, grattafiori2024llama, guo2025deepseek, jiang2024mixtral, granite2024granite} for reasoning, with tool calls. These agents also need to repeatedly interact with execution environments where the code can be executed and tested. 
As GitHub repositories and their required software dependencies evolve over time, library updates may cause failures when the agent interacts with a code base.
Because SWE agents are often complex, when the agent fails to solve its task, it is not clear if the failure was caused by a runtime error in an external component or the agent made a mistake. It is even harder to understand improvement or regression when running the agent with different hyper-parameters or when changing the underlying LLM.

An understanding of the inner workings of an agent can be helpful for debugging or for agent development.
Hence many agents \cite{wang2024openhands,xia2024agentless,moatless,yang2024swe} provide instrumentation to log critical steps of the agent run. 
However, these logs are often not ideal for human analysis and finding the relevant information can become time consuming. There is a need for tooling to consume the  information output by the agent into a visual format that is easy to understand by users and developers. 

We propose \systemname{}\footnote{We are actively looking to open source this software upon acceptance.}, a first of its kind, visualization framework designed to help SWE agent developers and researchers to acquire the relevant information regarding the agent's behavior. It is intended to aid in the development cycle of software agents: adding new capabilities to the agentic framework, LLM training for agentic tasks, hyper-parameter exploration.  \systemname{} is designed to analyze and provide insights from large scale experiments where the agent is asked to solve many instances of some SWE task e.g. GitHub issue resolution.

Overall, \systemname{} is the first AI agent visualizer that provides a ``one-stop shop'' for researchers with following capabilities to analyze AI Agents:

\begin{itemize}[leftmargin=*]
\item \textbf{Experiment Health}.  \systemname{} helps the developer validate that the run was executed as expected. It exposes the cause of failure including connectivity loss with the LLM, connectivity loss to the execution environment or agent mistakes.

\item \textbf{Experiment Comparison}.  \systemname{} provides insights on performance improvement or regression when changing hyper-parameters, LLMs or the tools provided to the agent. The experiment comparison facilitates the  development of new features including emerging approaches such as customization of Open-Source LLMs for the SWE task. 
This involves a series of experiments on a validation dataset to find the right hype-parameters, the correct checkpoint and the suitable training data mix. It is thus critical to be able to compare multiple experiments and select the best model. 

\item \textbf{Experiment Summarization}. \systemname{} supports summarization of multiple experiments for inference scaling, a test-time approach to improve the performance by repeatedly sampling the underlying LLM, increasing the chance to find the correct solution.

\item \textbf{Experiment Report}. \systemname{} also provides a performance report and a visualization of the instances that could not be evaluated due to malformed output.  While this information is easily available in the agent logs, for convenience we show it in a format that is easy to consume and understand. 

\end{itemize}

\noindent The \textbf{Experiment Health} and \textbf{Experiment Comparison} views have proven to be instrumental for the development of agents. We conducted a short user study and interviewed 10 researchers and developers with experience on SWE agents. All the users confirmed that the analysis is useful. 7 users reported that they had to collect the same information by writing bash commands and scripts. Users reported that understanding system health takes about 15 minutes for every experiment and comparing experiments can take between 10 and 30 minutes\footnote{10-30 minutes may not seem much for a single experiment, but aggregated over multiple runs it will save a lot of time for researchers working on this for a prolonged period.}.

\section{Related Work}
LLM agents use chain-of-thought \cite{wei2022chain}, tool calling \cite{berkeley-function-calling-leaderboard, schick2023toolformerlanguagemodelsteach}, self-debugging \cite{jimenez2024swebench}, sampling \cite{brown2024largelanguagemonkeysscaling, ehrlich2025codemonkeys} and search \cite{antoniades2024swe} strategies to iteratively solve GitHub issues as illustrated in SWE-Bench \cite{jimenez2024swebench}.  
SWE-Bench analysis~\cite{SweBenchAnalysis}, which is part of the SWE-Bench leaderboard, gives a limited high-level overview of the output by a single agent. A trajectory which is a sequence of thoughts and actions (via tools) taken by an agent, can offer useful insights on how it solves an issue and therefore can be used for debugging and improving the underlying LLM's performance. However, trajectories are usually quite long and visualization tools can be of great help in analyzing them.
There are a few tools like ~\citealp{OpenTelemetryHF},~\citealp{opentelemetryBeeAI} and~\citealp{langfuse} which allow individual trajectory level insights.
~\citealp{langfuse} also provides tool call related details at every step.
But they all stop short of high-level analysis or a way to compare multiple experiments. 
On the other hand, our proposed software, \systemname{}, provides researchers insights that includes experiment health, experiment report, experiment comparison and experiment summarization, thereby, providing a better debugging experience to engineers and researchers trying to improve model and agent performance on software engineering tasks. 
Recently, \citealp{desmond2025agent} provide a tool to visualize trajectories but, contrary to ours, they do not have software engineering oriented statistics or reports and they also do not have a way to compare experiments which is readily available via our proposed \systemname{}.

\section{\systemname{}} 
\subsection{Architecture}
\begin{figure}[]
  \includegraphics[width=\columnwidth]{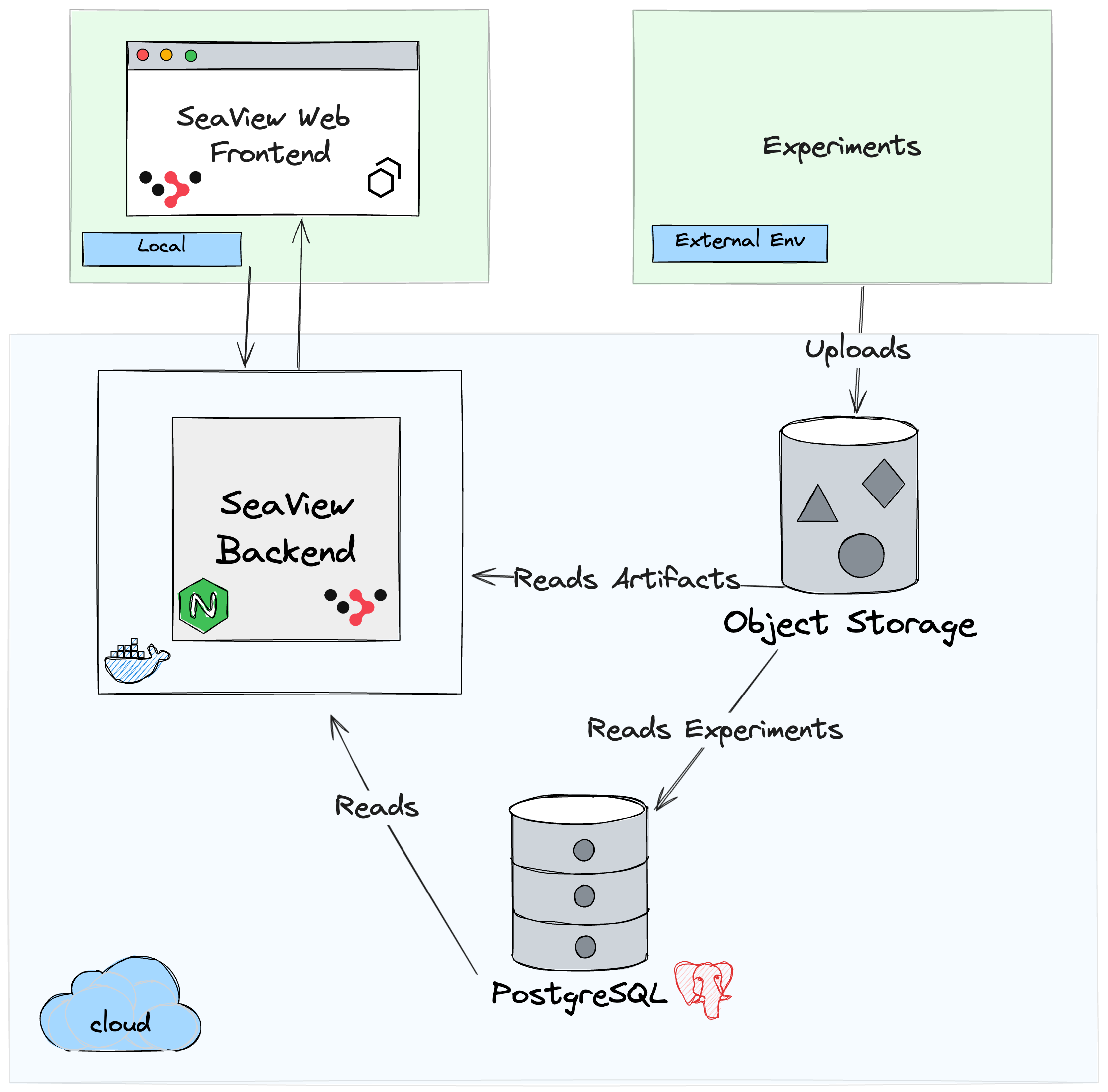}
  \caption{System Architecture}
  \label{fig:architecture}
\end{figure}

\systemname{} leverages a modern, full-stack \href{https://react.dev/learn/typescript}{TypeScript} system architecture. React Router v7, built with Node.js and React, serves as the full-stack web framework handling routing, data fetching, client-side interactivity and server-side operations. IBM’s open-source Carbon Design System \cite{carbondesign} is used to build the user interface. PostgreSQL provides a relational database for storing structured benchmark, framework and experiment data. Large files, such as raw benchmark outputs and logs, are stored in a cloud object store for scalable and cost-effective object storage. The application and Node.js runtime environment are containerized with Docker, facilitating portability and consistent deployment across various environments.

The experiment ETL pipeline as captured in Figure \ref{fig:architecture} shows how experiment data moves through the system, transformed from raw outputs to structured data and storage objects. Researchers run experiments in an external environment, for example, SWE-Bench Lite with the OpenHands agent framework. Once finished completed, the output directory is uploaded to a configured location in their object store bucket. Using a pull-based polling background job \systemname{} creates a record for experiment in the database and all of the associated records for the logs, instances and instance logs. Depending on the data, it is either stored directly in the database or references to the cloud object store are created. When analyzing experiments in the webapp, \systemname{} retrieves data from the database and the cloud object store.

\subsection{User Interaction}
\begin{figure*}[tb]
  \includegraphics[width=\linewidth]{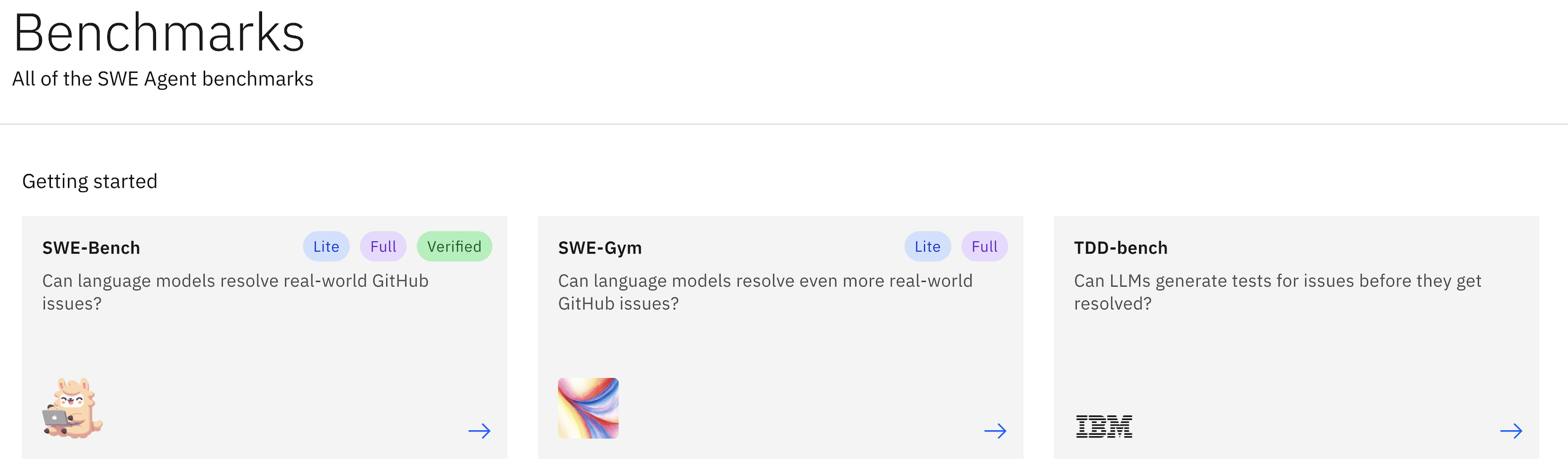} \hfill
  \includegraphics[width=\linewidth]{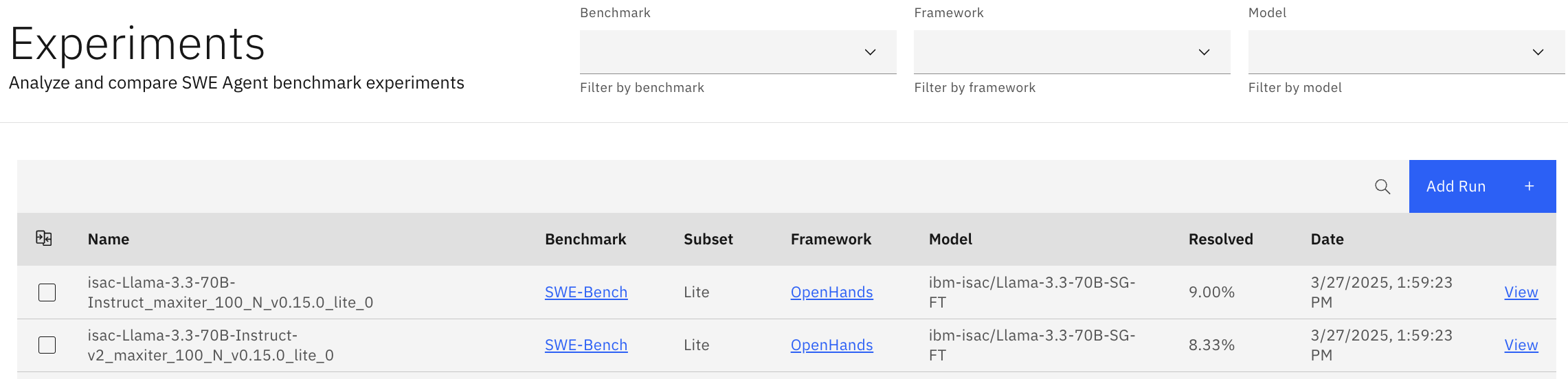} \hfill
  \caption{The Benchmarks View and the Experiments View.}
  \label{fig:benchmark_experiments}
\end{figure*}

\begin{figure*}[h]
  \includegraphics[width=0.3\linewidth]{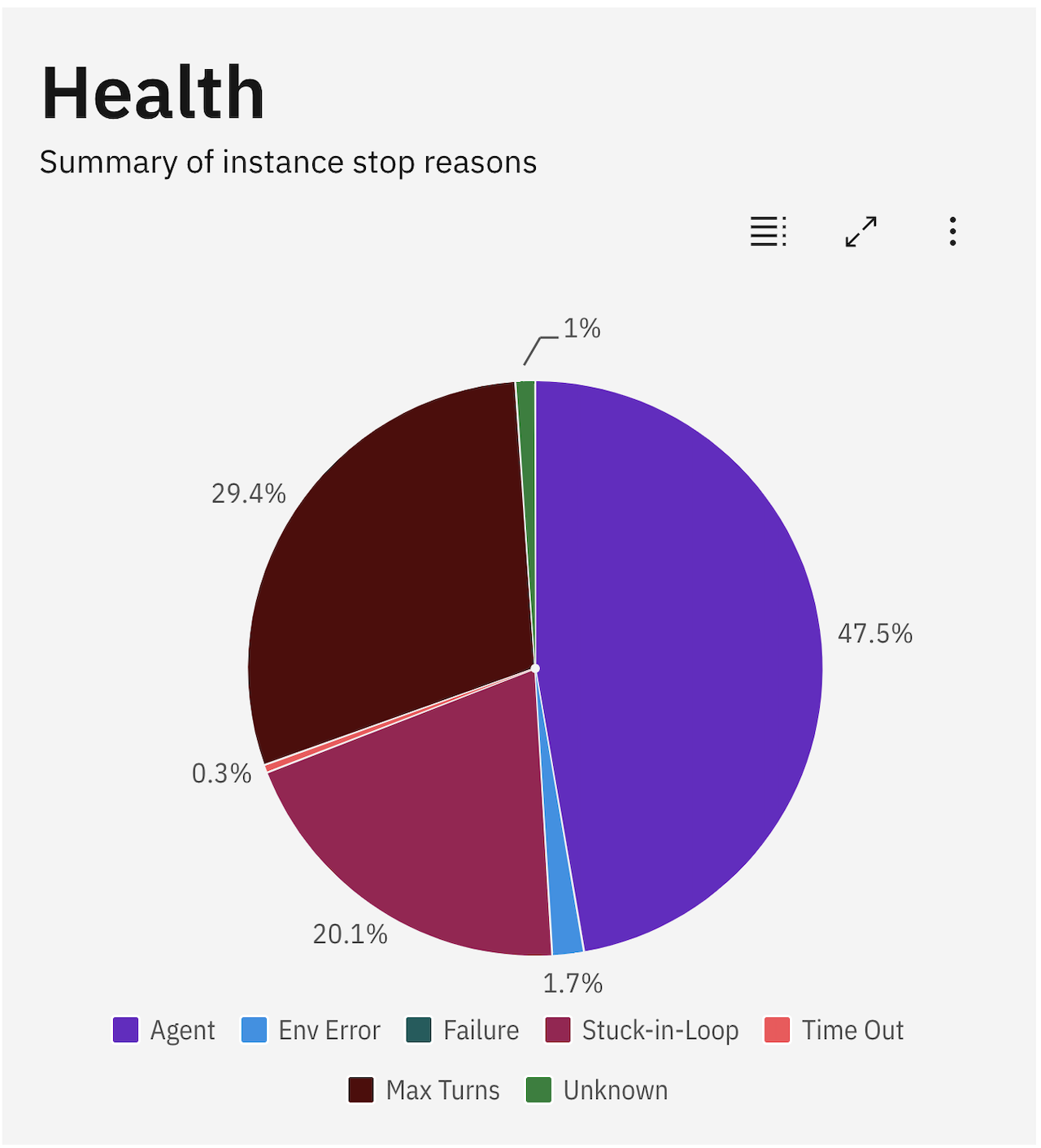} \hfill
  \includegraphics[width=0.7\linewidth]{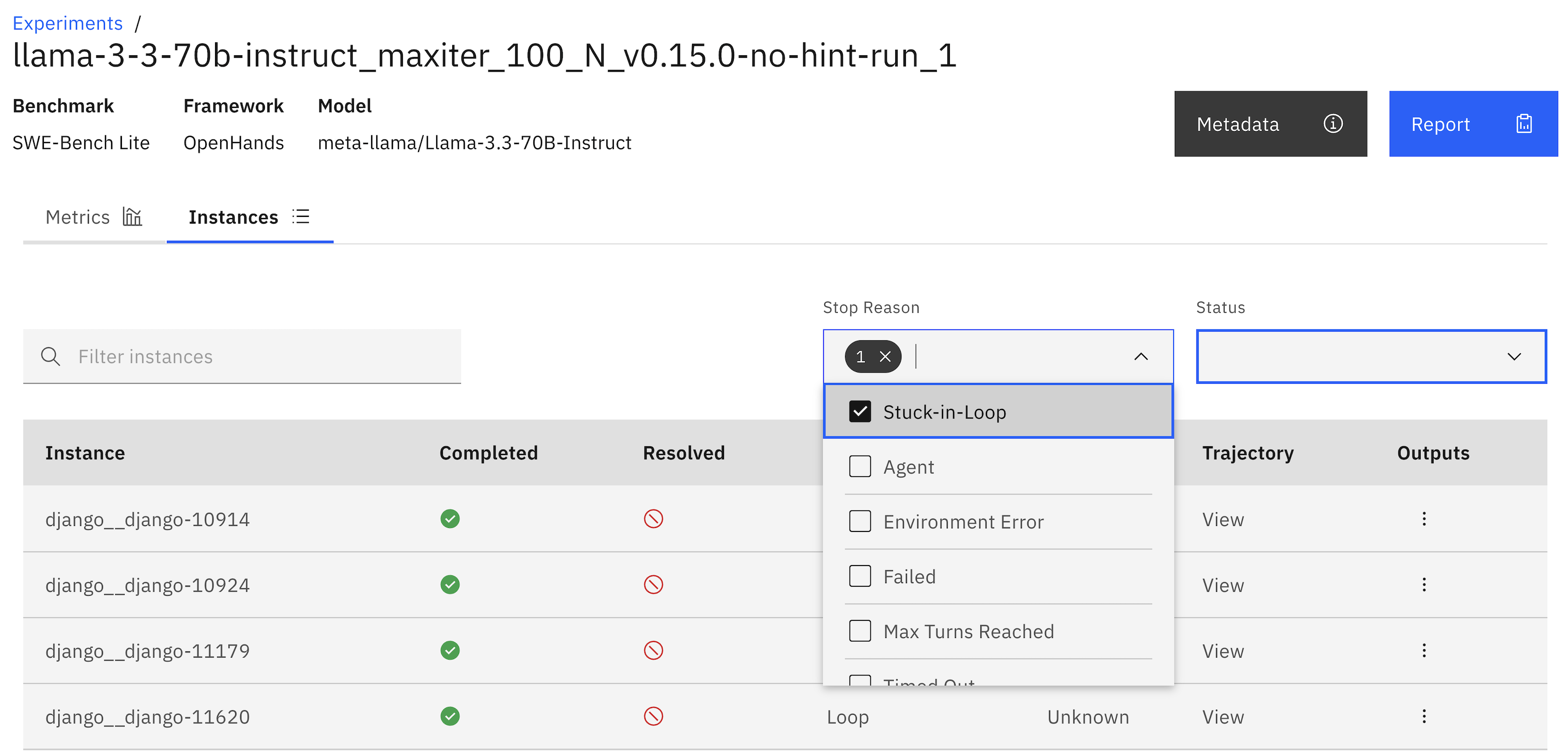}
  \caption {Left: The Experiment Health View showing the breakdown of the instance status for the entire experiment. Right: Instance View to filter and inspect instances using the same categories. }
  \label{fig:health_char_filter}
\end{figure*}

\begin{figure*}[h]
  \includegraphics[width=\linewidth]{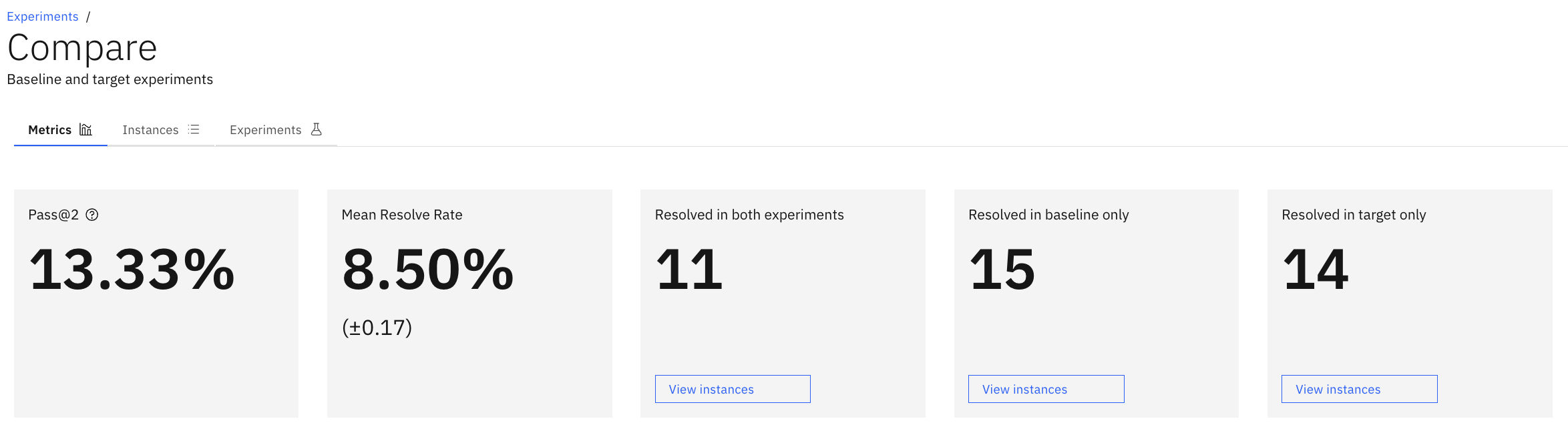} \hfill
  \includegraphics[width=\linewidth]{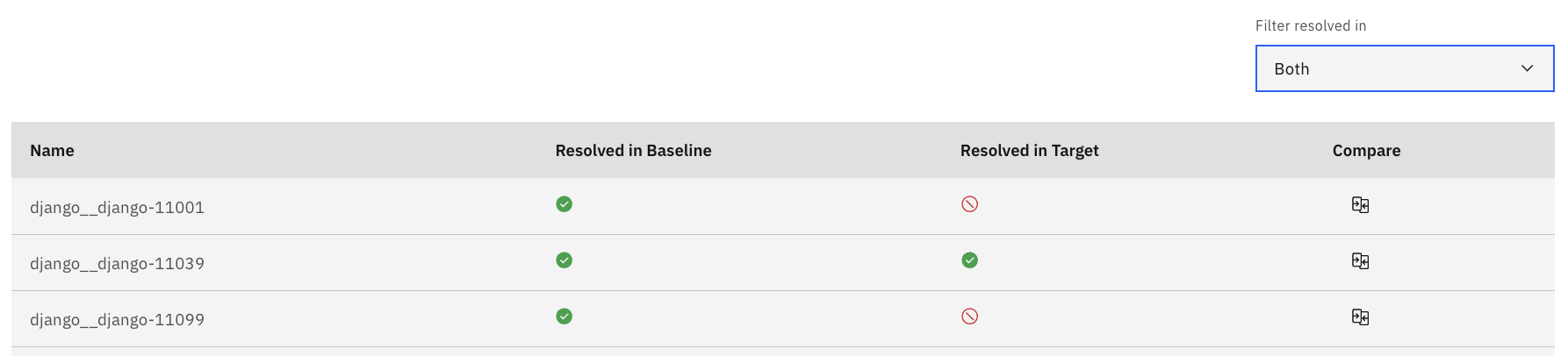} \hfill
  \caption{Top: Experiment Comparison View. Shows the number of instances resolved in each Experiment. Bottom: Instance view to analyze instance that are resolved in each experiment}
  \label{fig:experiment_compare}
\end{figure*}

\begin{figure*}[h]
  \includegraphics[width=\linewidth]{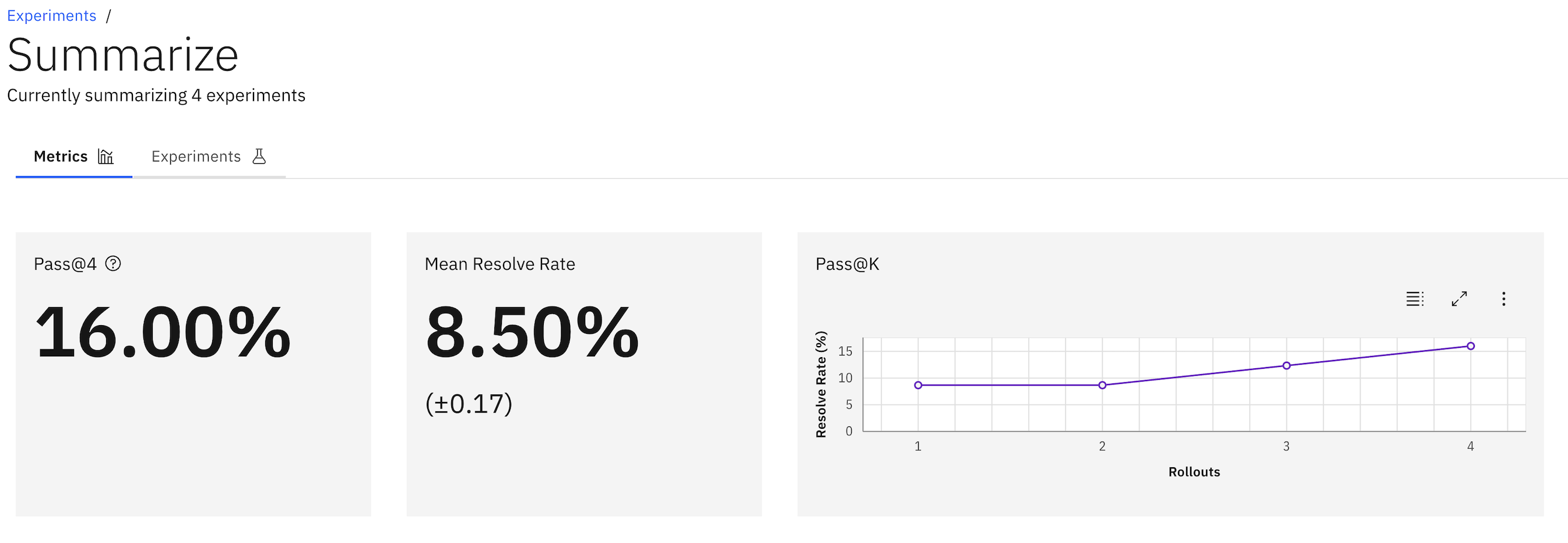} \hfill
  \caption{Experiment Summarization View. Shows the number of instances resolved by all selected experiments.}
  \label{fig:experiment_summary}
\end{figure*}

\systemname{} is a framework to analyze large scale agent experiments on datasets related to the software engineering tasks. Our main focus is on datasets for GitHub issue resolution including the SWE-Bench~\cite{SweBenchAnalysis} and the SWE-Gym~\cite{pan2024trainingsoftwareengineeringagents}. We are also planning support for related tasks such as unit test generation on the TDD-bench~\cite{Ahmed2024TDDBenchVC} dataset.
 
In this paper we present a common interaction with \systemname{} that is used when working with software engineering agents. In this scenario it is assumed that an agent developer has cloned an open source agent for the GitHub bug resolution, has followed the instructions to install the dependencies and has created the appropriate configuration file. The developer has run a first experiment on the SWE-Bench Lite dataset and the output of the experiment, produced by the agent, is uploaded into \systemname{}. 

The developer can now start the analysis, by selecting  SWE-Bench in the Benchmarks view provided by \systemname{} where all experiments that have been uploaded for the SWE-bench Lite task are available, as shown in Figure~\ref{fig:benchmark_experiments}.  Each experiment shows the agent framework, the underlying LLM and the percentage of resolved issues.  The developer selects the new experiment and can dive into the analysis. 

\textbf{Experiment Health} (Figure \ref{fig:health_char_filter}): This is the first and one of the most important phases of the analysis that gives insights on the root cause of agent failure. It exposes the instances that have failed due to issues in the experimental setup. These are instances where the agent can be successful if the failures are corrected. The developer can fix the environment setup and re-run the agent on the faulty instances. 
In the Instance analysis view the developer can see the status of every instance in the experiment and can select the instances of interest that have been identified through the Experiment Health.  For every instance, the framework exposes the GitHub issue description and the Git patch that was produced by the agent. The agent trajectory, containing all the steps the agent took to solve the problem, is also available.

\textbf{Experiment comparison} (Figure \ref{fig:experiment_compare}): This view is intended for researchers and developers who are making changes in the agent. In order to analyze the effect of these changes, the developers upload into \systemname{} their baseline experiment and a target experiment where the agent change is implemented. The Experiment Comparison view, depicted in Figure~\ref{fig:experiment_compare} shows the a breakdown of the instances indicating a performance gain. These are instances resolved in the target experiment but not in the baseline.
The view also provides the instances indicating a performance regression. The developer can select a particular instance from the validation dataset and compare the baseline and target trajectory, or the baseline and target patches. 

\textbf{Experiment summarization} (Figure \ref{fig:experiment_summary}): This view summarizes the outcome of multiple experiments. It is useful for inference-time scaling \cite{antoniades2024swe,pan2024trainingsoftwareengineeringagents} where the developer can easily determine the upper bound in agent performance and experiment variations when several solutions are sampled from the agent. 

\textbf{Experiment Report}. This view shows the performance report. SWE agents for GitHub issue resolution output a patch that resolves the bug. Instances for which the agent has produced an  Empty Patch, a Bad Patch or a patch that causes error during evaluation can be further inspected.

\section{Results}
We conducted a user study to understand how useful \systemname{} is for the AI/NLP/SWE researchers and developers. We enlisted 10 researchers and developers with practical experience in the SWE agent area. We described the Experiment Health and Experiment Comparison and asked question regarding their experience, as shown below.

\textit{
\textbf{\underline{Task Description:}} Based on your experience as software agent developer please answer the questions for the following scenarios:}

\textit{\textbf{\underline{Experiment Health:}} You run an experiment on the SWE-Bench Lite dataset and the task is to find which instances fail due to environment setup or agent configuration. These are instances that could be resolved by the agent if the experiment environment setup or the agent configuration are fixed.}

\textit{Examples of problems with the environment setup: LLM timeout, Docker container failures etc. Example of a problem with the agent configuration: the agent has exceeded the maximum number of turns. } 

\textit{\textbf{Question 1}: Is the task described above useful? (YES/NO) 
}

\textit{\textbf{Question 2}: Did you perform the task described above? (YES/NO) 
}

\textit{\textbf{Question 3}: If you performed the task, how long did it take on average for every experiment? }

\textit{\textbf{Question 4}: Did you have to create scripts or adjust your workflow to help gather this information?}

\textit{\textbf{\underline{Experiment Comparison:}} You are developing an agentic framework, and you are making a change in the agent: a new feature, a hyper-parameter change, different LLM or a new tool. You are comparing two experiments on the dev set: the baseline that does not include the change and the target experiment including the change. The task is to find instances where the change improved the agent and the instances where the change hurt the agent.}

\textit{\textbf{Question 1}: Is the task described above useful? (YES/NO)}

\textit{\textbf{Question 2}: Did you perform the task described above? (YES/NO)} 

\textit{\textbf{Question 3}: If you performed the task, how long did it take on average for every experiment? }

\textit{\textbf{Question 4}: Did you have to create scripts or adjust your workflow to help gather this information? }

For the Experiment Health view, 10/10 users confirmed that the information is useful. 8/10 users reported that they have performed the task to extract the Experiment Health information as part of their experimentation with agents. Extracting the relevant information took, on average, 15 minutes per experiment. Two users have reported that when they started experimenting with agents they needed between 30 minutes to 1 hour to investigate experiment health. 6/10 users reported they have written scripts to help them gather the Experiment Health data. 

For the Experiment Comparison view, 9/10 confirmed that the information is useful.  7/10 users reported that they have extracted the information as part of their experimentation with agents. Extracting the relevant information took between 10 and 30 minutes. Two users indicated that extracting the information for this scenario took significant amount of time but they are not able to provide an average per experiment. 5/10 users reported they have written scripts to help them gather the data.

\section{Conclusions}

We present \systemname{}, an analysis framework for large scale SWE agent experiments. 
This framework allows developers and researchers debug and improve agentic systems for SWE tasks.
To help analyze agents, \systemname{} provides capabilities like experiment health, experiment comparison and experiment summarization. 
An experiment report is also provided to show overall agent performance.

To enable the community understand the system better, we describe it's architecture and explain with screenshots the different capabilities and user interaction.
In order to validate the utility of our tool we conduct a user study with 10 researchers and developers working with SWE agents.
The questions in the study along with the results are shared and they indicate that the users found the tool useful and the analysis helped them save time.

This tool will help developers and researchers improve the performance of SWE agents. 
Upon acceptance, we plan to open source this framework so that the community can use it freely and develop it further.

\section*{Limitations}

\systemname{} is a great way to debug and improve SWE agents as it provides valuable insights at different levels of abstraction.
Still, there are some limitations in our tool which we hope to address in future versions.

\systemname{} helps analyze experiments which are already completed, but it is not a live dashboard of currently running experiments. 
Although \systemname{} already provides valuable information when comparing different SWE agent experiments, side-by-side comparison of generated trajectories and patches can be improved. 
Sampling, as part of inference scaling, is becoming an increasingly common component of agents. Currently, \systemname{} has limited support for sampling-specific visualization.

A tool call oriented comparison of experiments can help developers understand the interaction between the agent and its environment, test new features and develop new agentic capabilities.

\section*{Acknowledgments}

We would like to thank \citealp{desmond2025agent}, the authors of Agent Trajectory Explorer for inspiring this work.
The researchers and developers who participated in the user study helped us highlight the utility of \systemname{} and we are grateful for their feedback.
Watson Code Assistant~\cite{WCA} was used for writing some of the code for \systemname{}. 
No other AI assistant was used in the development of this work.

\section*{Ethics Statement}

\systemname{} by itself does not generate any non-ethical content (e.g. hate, abuse, profanity, etc) for researchers or developers. It does take as input LLM agent trajectories and hence are bound to the output of individual LLMs.

\bibliography{custom}

\begin{thebibliography}{28}
\providecommand{\natexlab}[1]{#1}

\bibitem[{Achiam et~al.(2023)Achiam, Adler, Agarwal, Ahmad, Akkaya, Aleman,
  Almeida, Altenschmidt, Altman, Anadkat et~al.}]{achiam2023gpt}
Josh Achiam, Steven Adler, Sandhini Agarwal, Lama Ahmad, Ilge Akkaya,
  Florencia~Leoni Aleman, Diogo Almeida, Janko Altenschmidt, Sam Altman,
  Shyamal Anadkat, and 1 others. 2023.
\newblock Gpt-4 technical report.
\newblock \emph{arXiv preprint arXiv:2303.08774}.

\bibitem[{Ahmed et~al.(2024)Ahmed, Hirzel, Pan, Shinnar, and
  Sinha}]{Ahmed2024TDDBenchVC}
Toufique Ahmed, Martin Hirzel, Rangeet Pan, Avraham Shinnar, and Saurabh Sinha.
  2024.
\newblock \href {https://api.semanticscholar.org/CorpusID:274464861} {Tdd-bench
  verified: Can llms generate tests for issues before they get resolved?}
\newblock \emph{ArXiv}, abs/2412.02883.

\bibitem[{Anthropic()}]{claude2}
Anthropic.
\newblock Claude 2.
\newblock \url{https://www.anthropic.com/}.

\bibitem[{Antoniades et~al.(2024)Antoniades, {\"O}rwall, Zhang, Xie, Goyal, and
  Wang}]{antoniades2024swe}
Antonis Antoniades, Albert {\"O}rwall, Kexun Zhang, Yuxi Xie, Anirudh Goyal,
  and William Wang. 2024.
\newblock Swe-search: Enhancing software agents with monte carlo tree search
  and iterative refinement.
\newblock \emph{arXiv preprint arXiv:2410.20285}.

\bibitem[{Brown et~al.(2024)Brown, Juravsky, Ehrlich, Clark, Le, Ré, and
  Mirhoseini}]{brown2024largelanguagemonkeysscaling}
Bradley Brown, Jordan Juravsky, Ryan Ehrlich, Ronald Clark, Quoc~V. Le,
  Christopher Ré, and Azalia Mirhoseini. 2024.
\newblock \href {https://arxiv.org/abs/2407.21787} {Large language monkeys:
  Scaling inference compute with repeated sampling}.

\bibitem[{Desmond et~al.(2025)Desmond, Lee, Ibrahim, Johnson, Sil, MacNair, and
  Puri}]{desmond2025agent}
Michael Desmond, Ja~Young Lee, Ibrahim Ibrahim, James Johnson, Avi Sil, Justin
  MacNair, and Ruchir Puri. 2025.
\newblock Agent trajectory explorer: Visualizing and providing feedback on
  agent trajectories.
\newblock In \emph{AAAI Conference on Artificial Intelligence}.

\bibitem[{Ehrlich et~al.(2025)Ehrlich, Brown, Juravsky, Clark, R{\'e}, and
  Mirhoseini}]{ehrlich2025codemonkeys}
Ryan Ehrlich, Bradley Brown, Jordan Juravsky, Ronald Clark, Christopher R{\'e},
  and Azalia Mirhoseini. 2025.
\newblock Codemonkeys: Scaling test-time compute for software engineering.
\newblock \emph{arXiv preprint arXiv:2501.14723}.

\bibitem[{Granite~Team(2024)}]{granite2024granite}
IBM Granite~Team. 2024.
\newblock \href {https://github.com/ibm-granite/granite-3.0-language-models/}
  {Granite 3.0 language models}.

\bibitem[{Grattafiori et~al.(2024)Grattafiori, Dubey, Jauhri, Pandey, Kadian,
  Al-Dahle, Letman, Mathur, Schelten, Vaughan et~al.}]{grattafiori2024llama}
Aaron Grattafiori, Abhimanyu Dubey, Abhinav Jauhri, Abhinav Pandey, Abhishek
  Kadian, Ahmad Al-Dahle, Aiesha Letman, Akhil Mathur, Alan Schelten, Alex
  Vaughan, and 1 others. 2024.
\newblock The llama 3 herd of models.
\newblock \emph{arXiv preprint arXiv:2407.21783}.

\bibitem[{Guo et~al.(2025)Guo, Yang, Zhang, Song, Zhang, Xu, Zhu, Ma, Wang, Bi
  et~al.}]{guo2025deepseek}
Daya Guo, Dejian Yang, Haowei Zhang, Junxiao Song, Ruoyu Zhang, Runxin Xu,
  Qihao Zhu, Shirong Ma, Peiyi Wang, Xiao Bi, and 1 others. 2025.
\newblock Deepseek-r1: Incentivizing reasoning capability in llms via
  reinforcement learning.
\newblock \emph{arXiv preprint arXiv:2501.12948}.

\bibitem[{IBM()}]{carbondesign}
IBM.
\newblock Carbon design system.
\newblock \url{https://carbondesignsystem.com/}.

\bibitem[{Jiang et~al.(2024)Jiang, Sablayrolles, Roux, Mensch, Savary, Bamford,
  Chaplot, Casas, Hanna, Bressand et~al.}]{jiang2024mixtral}
Albert~Q Jiang, Alexandre Sablayrolles, Antoine Roux, Arthur Mensch, Blanche
  Savary, Chris Bamford, Devendra~Singh Chaplot, Diego de~las Casas, Emma~Bou
  Hanna, Florian Bressand, and 1 others. 2024.
\newblock Mixtral of experts.
\newblock \emph{arXiv preprint arXiv:2401.04088}.

\bibitem[{Jimenez et~al.()Jimenez, Yang, Wettig, Yao, Pei, Press, and
  Narasimhan}]{SweBenchAnalysis}
Carlos~E Jimenez, John Yang, Alexander Wettig, Shunyu Yao, Kexin Pei, Ofir
  Press, and Karthik~R Narasimhan.
\newblock Swe-bench analysis.
\newblock \url{https://langfuse.com/}.

\bibitem[{Jimenez et~al.(2024)Jimenez, Yang, Wettig, Yao, Pei, Press, and
  Narasimhan}]{jimenez2024swebench}
Carlos~E Jimenez, John Yang, Alexander Wettig, Shunyu Yao, Kexin Pei, Ofir
  Press, and Karthik~R Narasimhan. 2024.
\newblock \href {https://openreview.net/forum?id=VTF8yNQM66} {{SWE}-bench: Can
  language models resolve real-world github issues?}
\newblock In \emph{The Twelfth International Conference on Learning
  Representations}.

\bibitem[{Langfuse()}]{langfuse}
Langfuse.
\newblock Langfuse.
\newblock \url{https://langfuse.com/}.

\bibitem[{OpenTelemetry-BeeAI()}]{opentelemetryBeeAI}
OpenTelemetry-BeeAI.
\newblock Opentelemetry instrumentation.
\newblock
  \url{https://i-am-bee.github.io/beeai-framework/#/python/instrumentation}.

\bibitem[{OpenTelemetry-HuggingFace()}]{OpenTelemetryHF}
OpenTelemetry-HuggingFace.
\newblock Inspecting runs with opentelemetry.
\newblock
  \url{https://huggingface.co/docs/smolagents/en/tutorials/inspect_runs}.

\bibitem[{Orwall()}]{moatless}
Albert Orwall.
\newblock Moatless tools.
\newblock \url{https://github.com/aorwall/moatless-tools}.

\bibitem[{Pan et~al.(2024)Pan, Wang, Neubig, Jaitly, Ji, Suhr, and
  Zhang}]{pan2024trainingsoftwareengineeringagents}
Jiayi Pan, Xingyao Wang, Graham Neubig, Navdeep Jaitly, Heng Ji, Alane Suhr,
  and Yizhe Zhang. 2024.
\newblock \href {https://arxiv.org/abs/2412.21139} {Training software
  engineering agents and verifiers with swe-gym}.

\bibitem[{Schick et~al.(2023)Schick, Dwivedi-Yu, Dessì, Raileanu, Lomeli,
  Zettlemoyer, Cancedda, and Scialom}]{schick2023toolformerlanguagemodelsteach}
Timo Schick, Jane Dwivedi-Yu, Roberto Dessì, Roberta Raileanu, Maria Lomeli,
  Luke Zettlemoyer, Nicola Cancedda, and Thomas Scialom. 2023.
\newblock \href {https://arxiv.org/abs/2302.04761} {Toolformer: Language models
  can teach themselves to use tools}.

\bibitem[{Wang et~al.(2024{\natexlab{a}})Wang, Li, Song, Xu, Tang, Zhuge, Pan,
  Song, Li, Singh et~al.}]{wang2024openhands}
Xingyao Wang, Boxuan Li, Yufan Song, Frank~F Xu, Xiangru Tang, Mingchen Zhuge,
  Jiayi Pan, Yueqi Song, Bowen Li, Jaskirat Singh, and 1 others.
  2024{\natexlab{a}}.
\newblock Openhands: An open platform for ai software developers as generalist
  agents.
\newblock In \emph{The Thirteenth International Conference on Learning
  Representations}.

\bibitem[{Wang et~al.(2024{\natexlab{b}})Wang, Chakravarthy, Munechika, and
  Chau}]{wang-etal-2024-wordflow}
Zijie Wang, Aishwarya Chakravarthy, David Munechika, and Duen~Horng Chau.
  2024{\natexlab{b}}.
\newblock \href {https://doi.org/10.18653/v1/2024.acl-demos.5} {Wordflow:
  Social prompt engineering for large language models}.
\newblock In \emph{Proceedings of the 62nd Annual Meeting of the Association
  for Computational Linguistics (Volume 3: System Demonstrations)}, pages
  42--50, Bangkok, Thailand. Association for Computational Linguistics.

\bibitem[{watsonx()}]{WCA}
watsonx.
\newblock Code smarter, not harder: watsonx code assistant.
\newblock \url{https://www.ibm.com/products/watsonx-code-assistant}.

\bibitem[{Wei et~al.(2022)Wei, Wang, Schuurmans, Bosma, Xia, Chi, Le, Zhou
  et~al.}]{wei2022chain}
Jason Wei, Xuezhi Wang, Dale Schuurmans, Maarten Bosma, Fei Xia, Ed~Chi, Quoc~V
  Le, Denny Zhou, and 1 others. 2022.
\newblock Chain-of-thought prompting elicits reasoning in large language
  models.
\newblock \emph{Advances in neural information processing systems},
  35:24824--24837.

\bibitem[{Xia et~al.(2024)Xia, Deng, Dunn, and Zhang}]{xia2024agentless}
Chunqiu~Steven Xia, Yinlin Deng, Soren Dunn, and Lingming Zhang. 2024.
\newblock Agentless: Demystifying llm-based software engineering agents.
\newblock \emph{arXiv preprint arXiv:2407.01489}.

\bibitem[{Yan et~al.(2024)Yan, Mao, Ji, Zhang, Patil, Stoica, and
  Gonzalez}]{berkeley-function-calling-leaderboard}
Fanjia Yan, Huanzhi Mao, Charlie Cheng-Jie Ji, Tianjun Zhang, Shishir~G. Patil,
  Ion Stoica, and Joseph~E. Gonzalez. 2024.
\newblock Berkeley function calling leaderboard.
\newblock
  \url{https://gorilla.cs.berkeley.edu/blogs/8_berkeley_function_calling_leaderboard.html}.

\bibitem[{Yang et~al.(2024)Yang, Jimenez, Wettig, Lieret, Yao, Narasimhan, and
  Press}]{yang2024swe}
John Yang, Carlos Jimenez, Alexander Wettig, Kilian Lieret, Shunyu Yao, Karthik
  Narasimhan, and Ofir Press. 2024.
\newblock Swe-agent: Agent-computer interfaces enable automated software
  engineering.
\newblock \emph{Advances in Neural Information Processing Systems},
  37:50528--50652.

\bibitem[{Zhao et~al.(2024)Zhao, Jiang, Lee, Chiu, Cardie, Gall{\'e}, and
  Rush}]{zhao2024commit0}
Wenting Zhao, Nan Jiang, Celine Lee, Justin~T Chiu, Claire Cardie, Matthias
  Gall{\'e}, and Alexander~M Rush. 2024.
\newblock Commit0: Library generation from scratch.
\newblock \emph{arXiv preprint arXiv:2412.01769}.

\end{thebibliography}

\appendix



\end{document}